\documentclass[published]{JHEP3}

\usepackage{epsfig}

\skip\footins = 1\bigskipamount plus 2pt minus 4pt

\newcommand{\Tr}{\mathop{\rm Tr}\nolimits}
\newcommand{\sfrac}[2]{{\frac{#1}{#2}}}

\title{Recombination of intersecting D-branes and cosmological inflation}

\author{Marta G\'omez-Reino and Ivonne Zavala C.\\
Centre for Mathematical Sciences, DAMTP, University of Cambridge\\
Cambridge, CB3 0WA, UK\\
E-mail: \email{mgr25@damtp.cam.ac.uk}, \email{eiz20@damtp.cam.ac.uk}}

\abstract{We consider the interactions between D$p$-branes
intersecting at an arbitrary number of angles in the context of
type-II string theory. For cosmology purposes we concentrate in the
theory on $\mathbb{R}^{3,1} \times T^6$. Interpreting the distance
between the branes as the inflaton field, the branes can intersect at
most at two angles on the compact space. If the configuration is non
supersymmetric we will have an interbrane potential that provides an
effective cosmological inflationary epoch at the four-dimensional
intersection between the branes.  The end of inflation occurs when the
interbrane distance becomes small compared with the string scale,
where a tachyon develops triggering the recombination of the
branes. We study this recombination due to tachyon instabilities and
we find the possibility for the final configuration to be again branes
intersecting at two angles. This preserves the interesting features
that are present in the intersecting brane models from the string
model building point of view also after the end of inflation. This
fact was not present in the models of branes intersecting at just one
angle. This kind of recombination can be also important in other
string contexts. \vskip0.4cm \hfill\break
\leftline{DAMTP-2002-95,  hep-th/0207278}}


\received{August 14, 2002}

\accepted{September 06, 2002}

\JHEP{09(2002)020}

\begin{document}

\section{Introduction}

Inflationary cosmology~\cite{guth} continues being the most likely
scenario that provides an explanation to many of the problems in
standard cosmology, such as the flatness, the horizon, or the
monopole problems. However, it is still a successful idea seeking
for a successful theoretical realization. In this sense, it is
natural to look for such a theoretical realization within string
theory, since it can provide us with a quantum description of
gravity at high energies. In  fact,  there has been recently a
lot of work towards a consistent realization of  this scenario
using different brane configurations which appear naturally in
string theory~\cite{tye}--\cite{st}. In these models the inflaton
potential has been  identified with one of the moduli potentials
either coming from the interactions between the branes or from
the complex structure of the compact space.

Systems with interacting branes in  non-supersymmetric
configurations seem to be natural candidates for inflation. One
of the reasons for this is that their interactions are well known
within string theory (see for example \cite{pol}) so one can
calculate the attractive potential between the branes. Another
reason is that, since  the configurations  are  unstable --- as
they are non supersymmetric --- tachyons will in general appear
opening the possibility of a hybrid inflationary~\cite{linde}
regime which would  naturally trigger the end of inflation. This
possibility was first described in~\cite{quei}.

Of particular interest are the models with brane configurations
where the supersymmetry is only slightly
broken~\cite{kali,shafi,kalii,g-b}. In particular, within this
class of models, configurations with branes intersecting at
angles are specially attractive. The main reason \pagebreak[3] is
that chiral fermions appear naturally at  the intersections
between the branes~\cite{douglas}. This property has been
exploited widely in the construction of (semi) realistic brane
models  with a spectrum very close~\cite{lustii}--\cite{koko}  to
that of the standard model  of particle physics. There are also
some supersymmetric versions of these intersecting brane
models~\cite{uri}.

Within the context of intersecting brane models, in
ref.\cite{g-b} it was shown that it is possible to obtain a
successful  inflationary period from slightly non-supersymmetric
configurations consisting of  D4-branes intersecting at one angle
on a six-dimensional compact space. Since  the supersymmetry was
broken that gave rise to an interaction potential between the
branes parametrized by the interbrane distance, which was
identified with the inflaton field. Inflation ended when the
distance between the branes became of order of the string scale.
At that moment a tachyon appeared triggering the recombination of
the branes  to a single brane stable configuration, and then
losing  the chirality property at  the intersection present at
the beginning of inflation.

\looseness=-1
Following this line, we first consider the most general
configuration of branes making $n$ angles in type-II string
theory. We then concentrate on the theory on $\mathbb{R}^{3,1}
\times T^6$. In that case we study the most general angled
configuration that allows for an interbrane potential in terms of
the separation between the branes --- on the compact space --- and
that can give rise to an inflationary period due to this
interbrane distance (in the effective four-dimensional theory).
As we will see, this requirement implies that  the maximal number
of angles that the branes can make on the compact space is $n=2$.
As will be explained below, a successful inflationary epoch will
occur as long as the starting configuration is not far away from
the supersymmetric one. An interesting feature of these
configurations, compared with the one  angled D-brane case, is
the fact that branes intersecting at two angles in the compact
space open the  possibility of ending up, after the recombination
of the branes (that is, after the end of inflation), with a final
state being a supersymmetric configuration with again two branes
intersecting at angles. This is the most  attractive property of
these configurations since they would provide a first step
towards the construction of semi realistic (supersymmetric) brane
models with chiral matter after the end of inflation in the
context of intersecting  brane models (the possibility of
obtaining a realistic model after inflation was already
considered in~\cite{queii} in the context of brane-antibrane
interactions at orbifold singularities). This possibility was not
present in the case of branes intersecting at just one angle.

The structure of the paper is as  follows: in section~\ref{sec:2}
we explain the general type-II string brane interactions between
branes at angles. Then we concentrate in the two angled case and
at the end we discuss the different possibilities for the
recombination of the system that appear in this case. In
section~\ref{sec:3} we  review the basic inflationary parameters
and concentrate in the two specific brane models which can give
rise to a successful inflationary epoch. In section~\ref{sec:4}
we  give some comments about the end of inflation and in the last
section we will conclude.

\section{The string model}\label{sec:2}

This section is divided in three parts. In the first one, we
review the interaction potential that appears between D$p$-branes
intersecting each other at different angles. In the second
subsection we explain the string model that we are interested in,
and also the approximations that we will consider for a suitable
application to inflation. Finally, in the last subsection we
analyse the decay and recombination of the system due to tachyon
instabilities.

\subsection{Interactions between intersecting D-branes}

Let us consider two D$p$-branes in type-II string theory,
intersecting at $n$ angles inside the ten-dimensional space. We
will review in this subsection the interaction potential arising
when the two D$p$-branes are located in such a way that they
intersect at a number $n$ of angles, $\Delta\theta_1,\dots
\Delta\theta_n$, being the remaining $(p+1-n)$ dimensions of the
branes either parallel to each other or common to both branes.

The interaction between the branes can be computed from the
exchange of massless closed string modes. This can be computed
from the one-loop vacuum amplitude for the open strings stretched
between the two D$p$-branes, that is given by
\begin{equation}
\mathcal A = 2 \int\frac{dt}{2t}\Tr  e^{-tH}\,,
\end{equation}
where $H$ is the open string hamiltonian. For two D$p$-branes
making  $n$ angles in ten dimensions  this amplitude can be
computed to give~\cite{jab,pol}\footnote{Configurations with more
than two branes intersecting with each other at angles, in only
one plane, were considered in~\cite{vancea}.}
\begin{equation}\label{amplitud}
\mathcal A = V_p\int_0^\infty\frac{dt}{t}
\exp^{-\frac{t\,Y^2}{2\pi^2\alpha'}}
(8\pi^2\alpha't)^{-\frac{p-3}{2}}
\left(-iL\eta(i\,t)^{-3}(8\pi\alpha't)^{-{1}/{2}}\right)^{4-n}
(Z_{NS}-Z_{R})\,,
\end{equation}
with
\begin{eqnarray}
Z_{NS} &=& (\Theta_3 (0\mid it))^{4-n}
\prod_{j=1}^n\frac{\Theta_3(i \Delta\theta_j t\mid it)}
{\Theta_1(i \Delta\theta_j t \mid it)} -
(\Theta_4 (0\mid it))^{4-n}
\prod_{j=1}^n\frac{\Theta_4(i \Delta\theta_jt \mid it)}
{\Theta_1(i \Delta\theta_j t \mid it)}\,, \nonumber\\
Z_{R} &=&(\Theta_2 (0\mid it))^{4-n}
\prod_{j=1}^n\frac{\Theta_2 (i\Delta\theta_j t \mid it)}{
\Theta_1(i \Delta\theta_jt \mid it)}\,,\label{zeta}
\end{eqnarray}
being the contributions coming from the $NS$ and $R$ sectors.
Also in~(\ref{zeta}) $\Theta_i$ are the usual Jacobi functions
and $\eta$ is the Dedekin function. Furthermore,
in~(\ref{amplitud}) by $Y$ we mean the distance between both
branes, $Y=\sqrt{\sum_k Y_k^2}$, where $k$ labels the dimensions
in which the branes are separated and $Y_k$ the distance between
both branes along the $k$ direction.

The various terms in the expression~(\ref{amplitud}) can be understood
as follows: a general D$p$-brane in ten dimensions can make at most
$\min(p,9-p)$ angles, e.g.\ a D3-brane can make at most 3 angles, and
a D7-brane can make at most 2 angles, etc.  Each time that one angle
is taken to zero, a factor of
$-iL\eta(i\,t)^{-3}(8\pi\alpha't)^{-{1}/{2}}$ appears in the
amplitude, reflecting the fact that the branes become parallel in one
dimension, where $V_pL^{4-n}$ gives the volume of the common
dimensions to both branes.

We are interested now in the  small $t$ limit
of~(\ref{amplitud}), that is, the large distance limit ($Y\gg
l_s$). This is the right limit that takes into account the
contributions coming from the massless closed strings exchanged
between the branes. Using the well known modular properties of
the $\Theta$ and $\eta$ functions we obtain, in the $t
\rightarrow 0$ limit, that the amplitude is just given by
\begin{equation}
\mathcal A(Y,\Delta\theta_j)= \frac{V_p L^{4-n}F(\Delta
\theta_j)}{2^{p-2}(2\pi^2\alpha')^{(p+1-n)/2}}
\int t^{-\frac{p+n-5}{2}}\exp^{-\frac{t\,Y^2}{2\pi^2\alpha'}}dt\,,
\label{int}
\end{equation}
where the function $F$ contains  the dependence on the relative
angles between the branes, and is extracted from the small $t$
limit of~(\ref{zeta}). The exact form of this function is given by
\begin{equation}
F(\Delta\theta_j)=\frac{(4-n)+\sum_{j+1}^n\cos2\Delta\theta_j
-4\prod_{j=1}^n \cos\Delta\theta_j}{2\prod_{j=1}^n
\sin\Delta\theta_j}\,.
\label{ef}
\end{equation}
The relation between the angles $\Delta\theta_j$  that make this
function $F$ vanish, correspond to the supersymmetric
configurations of the two brane system for each given number of
angles. For example it is easy to see that for branes at just one
angle the function~(\ref{ef}) only vanishes when
$\Delta\theta_1=0$,  corresponding to the supersymmetric
situation. As we will see later, in the two angled case the
function~(\ref{ef}) vanishes when
$\Delta\theta_1=\pm\Delta\theta_2$~\cite{jab}.

The interaction potential between the branes can then be
calculated by performing the integral~(\ref{int}). This integral
is just given in terms of the Euler $\Gamma$-function, so the
potential has the following form
\begin{equation}\label{amplitude}
V(Y,\Delta\theta_j)= -\frac{V_pL^{4-n}F(\Delta\theta_j)}
{2^{p-2}(2\pi^2\alpha')^{p-3}}\Gamma
\left(\frac{7-p-n}{2}\right)Y^{(p+n-7)}\,.
\end{equation}
Note that for $p+n=7$ this expression is not valid as $\Gamma(0)$
is not a well defined function. In fact in that case the
integral~(\ref{int}) is divergent, so we need to introduce a
lower cutoff to perform it. If we denote by $\Lambda_c$ the
cutoff, the integral becomes
\begin{equation}\label{amplitudedos}
V(Y,\Delta\theta_j)= \frac{V_pL^{p-3}F(\Delta\theta_j)}
{(4\pi^2\alpha')^{p-3}} \ln\frac{Y}{\Lambda_c}\,.
\end{equation}

Taking a quick look at  the form of the
potentials~(\ref{amplitude}),~(\ref{amplitudedos}), it is natural
to find them interesting from the cosmological point of view. In
fact if one identifies the distance $Y$ between the branes as the
inflaton field, it is clear that the potential can be made flat
enough  by choosing appropriately the angles. This can be seen
straight forwardly by noticing that the function $F$ can be made
very small as one approaches to a supersymmetric configuration.
Also it is interesting to check that the
potentials~(\ref{amplitude}),~(\ref{amplitudedos}) are always
attractive since the $\Gamma$-function changes  sign when its
argument becomes negative.

Now, we are interested  in inflationary cosmology in four
dimensions coming from this configurations, so we  concentrate in
models where both branes span  four non-compact dimensions and
intersect over the remaining six dimensions  that we  take to be
compactified. We then want to identify the inflationary field
with the interbrane distance on the compact six-dimensional space.
Since both branes are extended and parallel in the four
non-compact dimensions, the maximum number of angles that they can
make, having an interbrane distance different from zero, is two.
That is, the most general situation to consider in such
configurations is take $n=2$ in~(\ref{amplitude}). The situation
with just one angle can then be extracted from the two angled
case. In fact that situation was considered in~\cite{g-b,st} so
we will concentrate here only in the case of branes intersecting
at two angles. The precise model under study will be discussed in
detail in the following subsection.

\subsection{Description of the model}
Let us then consider now two D$p$-branes in type-II string theory
spanning $(3+1)$ dimensions along the non-compact space
$\mathbb{R}^{3,1}$, with their remaining spatial dimensions
laying on the compact space, that  is taken to be a factorisable
six torus, i.e.\ $T^6=T^2_{(1)}\times T^2_{(2)}\times T^2_{(3)}$.
As explained above, since we want to have a non-zero interbrane
separation $Y$,  the most general possibility that we can
consider is take D5 or D6-branes making two angles on the compact
space. We  discuss now this situation.

As it is shown in figure~\ref{fig:1} we will consider that the
branes wrap some factorisable 2-cycles along two of the tori of
the compact space, $T^2_{(1)}\times T^2_{(2)}$, and  they are
separated by a distance $Y=\sqrt{\sum_k Y_k^2}$ on the third
torus $T^2_{(3)}$. We will assume for simplicity  that we have
squared tori with radii $R$, and that each D$p_i$-brane wraps a
factorisable cycle with homological charge
$(n^{(i)}_I,m^{(i)}_I)$ on the $I'$-th torus. This means that each
brane wraps a number $(n^{(i)}_I,m^{(i)}_I)$ of times over the
corresponding homology cycle of the $I'$-th torus.
\EPSFIGURE[t]{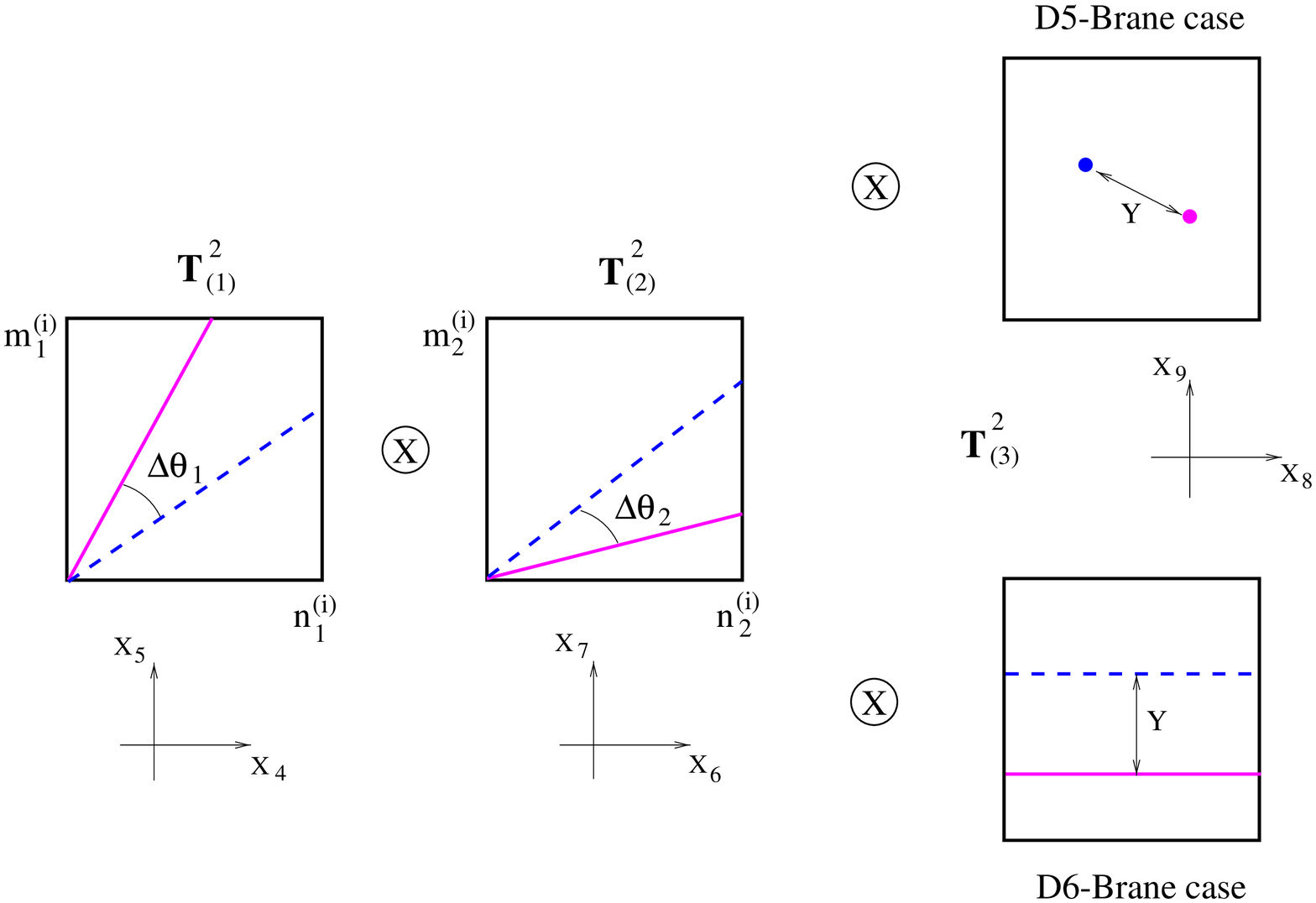, width=.826\textwidth}
{Diagram of the configurations we are considering in the two
cases discussed.\label{fig:1}}

The angle between the branes for squared tori, as can be seen in
figure~\ref{fig:1}, is given in terms of the homological charges
of the branes on each torus by the formula
\begin{equation}
\theta_I^{(i)}=\mbox{arctan}\frac{m_{I}^{(i)}}{n_{I}^{(i)}}\,,\qquad
I=1,2\,,\qquad \Delta\theta_I=\left|\theta_I^{(1)}-\theta_I^{(2)}\right|\,.
\label{angle}
\end{equation}

Note that if both branes are wrapped on the same cycles and with
the same orientations in each tori (i.e.\
$\Delta\theta_1=\Delta\theta_2=0$ or $\Delta\theta_1=
\Delta\theta_2=\pi$) that means that they are parallel. In this
case we have an $\mathcal{N}=4$ supersymmetric theory from the
four-dimensional point of view, that is, we have a configuration
that preserves sixteen supercharges. If they wrap the same cycles
but with opposite orientations along one of the two tori and the
same orientation on the other (i.e.\ $\Delta
\theta_1=0, \Delta\theta_2=\pi$ or vice versa) we have a
brane-antibrane configuration, which breaks completely the
supersymmetry.

In a generic case with topologically different cycles we will
have a configuration where the branes intersect at non-zero
relative angles $\Delta\theta_I$  in the range $0 \leq
\Delta\theta_I \leq \pi$, $I=1,2$. This configuration is
supersymmetric when $|\Delta \theta_1 | = |\Delta
\theta_2 |$, and it preserves $1/4$ of the supersymmetries
(that is, we will have a $\mathcal{N}=1$ supersymmetric theory
from the four-dimensional point of view)~\cite{jab}. For this
configuration the total homology class of the $2$-cycles is the
sum of the homology classes of the two branes, being --- in the
supersymmetric case --- a special lagrangian in the complex
structure. In the general case, that is, when $|\Delta \theta_1 |
\neq |\Delta \theta_2 |$, all the supersymmetries are broken.

In this non-supersymmetric case (considering that
$\Delta\theta_I\in[0,\pi]$) the mass of the lightest scalars is
controlled by the angles between the D$p$-branes and is given
by~\cite{douglas,jab}
\begin{eqnarray}
\alpha'M^2_1 &=& \frac{Y^2}{4\pi^2\alpha'}+\frac12(\Delta\theta_1-
\Delta\theta_2)\,,\nonumber\\
\alpha'M^2_2 &=& \frac{Y^2}{4\pi^2\alpha'}+\frac12(\Delta\theta_2-
\Delta\theta_1)\,,\nonumber\\
\alpha'M^2_3 &=& \frac{Y^2}{4\pi^2\alpha'}+\frac12(\Delta\theta_1+
\Delta\theta_2)\,,\nonumber\\
\alpha'M^2_4 &=& \frac{Y^2}{4\pi^2\alpha'}+1-\frac12(\Delta\theta_1+
\Delta\theta_2)\,,\label{mdos}
\end{eqnarray}
where $Y$ denotes the separation between the branes in the third
torus. This indicates that in any non-supersymmetric case one of
these complex scalars will become tachyonic.

The tachyon that appears at the intersection will trigger the
recombination of the branes. In the general case, the branes will
recombine to only  one brane wrapping the minimal volume
supersymmetric 2-cycle in the total homology class~\cite{uranga}.
Nevertheless, as we will explain  in the last part of this
section, there are some  cases in which the final state can  be a
supersymmetric configuration with still two branes intersecting
at angles.

Now that we have explained the general features of the model we
have in mind,  we must point out some relevant considerations
that should be done about the model: first of all note that the
R-R vanishing tadpole conditions must be satisfied in order for
the theory to be well defined. When we have branes intersecting
at angles on a compactified space, the R-R tadpole vanishing
conditions are equivalent to the vanishing of the total homology
charge of the system. In our configuration this is only fulfilled
in the special case that we have a brane and an antibrane. In the
general case we will need an extra brane with opposite total
homology charge for the conditions  to be satisfied. However this
is not important for our analysis as we can place an additional
brane at a large distance in the transverse directions, so it
would not be involved in the dynamical evolution of the above
mentioned brane configuration.

Another important assumption that we will make  along this paper
is that the moduli associated to the  compact space  and the
dilaton are stabilized by some unknown mechanism. In other words,
we will consider that the evolution of the closed string modes is
much slower than the evolution of the open string modes. That
means that we will only consider the attractive potential between
the branes due to the interbrane distance $Y$ as the relevant
potential, which will be identified with the inflaton potential.
This is the strongest assumption we will make and it is consistent
as soon as the overall size of the compact space is larger
compared with the distance between the branes.

Apart from the contribution coming from the R-R tadpole
conditions we have to consider also the contribution coming from
the NS-NS tadpoles. When a brane configuration is supersymmetric
we have (from the closed string modes) that the contribution due
to the NS-NS tadpoles (those that came from the dilaton plus
graviton interactions) cancel exactly the contribution coming
from the R-R tadpoles. If we break the supersymmetry of the system
we will have that the attractive potential coming from the
dilaton and graviton interactions will be greater than the
repulsive potential due to the fact that both branes have same
sign R-R charge. As a result, the branes will attract to each
other. Then, in this case, since the supersymmetry of the system
is being broken, there are going to be uncancelled NS-NS tadpoles
that will act as a potential between the branes. This potential
has been calculated in the previous subsection for the most
general case, on a non-compact space. When dealing with compact
spaces the expression~(\ref{int}) is modified in the following way
\begin{equation}
\mathcal{A}(Y,\Delta\theta_j)= \frac{V_p
L^{4-n}F(\Delta\theta_j)}{2^{p-2}(2\pi^2\alpha')^{(p+1-n)/2}}
\sum_{\omega_k\in Z} \int_0^\infty  t^{-\frac{p+n-5}{2}}
\exp^{-\frac{t\sum_k(Y_k+2\pi \omega_k R)^2}
{2\pi^2\alpha'}}dt\,,
\label{intdos}
\end{equation}
where $\omega_k$ represents the winding modes of the strings on
the directions transverse to the branes.\footnote{That means that
the summation over $k$ in~(\ref{intdos}) has only one term in the
D$6$-brane case and it will be $Y_9=|x_9^{(1)}-x_9^{(2)}|$ (see
figure~\ref{fig:1}). In the D$5$-brane case we will have two
terms: $Y_8=|x_8^{(1)}-x_8^{(2)}|$ and
$Y_9=|x_9^{(1)}-x_9^{(2)}|$. Also in both cases we will denote
$Y=\sqrt{\sum_k Y_k^2}$.} Nevertheless, if the distance between
the branes is small compared with the compactification radii
($Y\ll(2\pi R)$), the winding modes would be too massive and
then  will not contribute to the low energy regime. That is, it
will cost a lot of energy to the strings to wind around the
compact space. If we translate this assumption to~(\ref{intdos}),
the dominant mode will be the zero mode, and the potential can be
written as in~(\ref{amplitude}),~(\ref{amplitudedos}), taking
into account that we focus on the case where the number of angles
is $n=2$. In this case the potential (when normalised over the
non-compact directions) for branes of different dimensions is just
given by
\begin{eqnarray}
V_{Dp}(Y,\Delta\theta_j)&=& -\frac{(2\pi R)^{(p-5)}F(\Delta\theta_j)}
{2^{p-2}(2\pi^2\alpha')^{p-3}} \Gamma
\left(\frac{5-p}{2}\right)Y^{(p-5)}\,,\label{pots}\\
V_{D5}(Y,\Delta\theta_j)&=& \frac{F(\Delta\theta_j)}
{(4\pi^2\alpha')^{2}} \ln\frac{Y}{\Lambda_c}\,,\label{potf}
\end{eqnarray}
where the $(2\pi R)^{p-5}$ factor arises from the dimensions in which
the branes become parallel on the compact dimensions (remember that
$R$ denotes the radius of the torus).

In the case of a two angled configuration the function
$F(\Delta\theta_j)$ given in~(\ref{ef}) can be written as
\begin{equation}
F(\Delta\theta_1,\Delta\theta_2)=\frac{\left(\cos\Delta\theta_1
-\cos\Delta\theta_2\right)^2}{\sin\Delta\theta_1
\sin\Delta\theta_2}\,.
\label{efe}
\end{equation}
Note that this expression has the right behaviour in the
supersymmetric cases since it vanishes when $\Delta\theta_1=
\pm\Delta\theta_2$, as expected. In the limit
$\Delta\theta_1=\pm\Delta\theta_2+\Delta\theta$, where the
supersymmetry is slightly broken, that is $\Delta\theta\ll 1$, we
find that $F(\Delta\theta_1, \Delta\theta_2)\simeq
(\Delta\theta)^2$.

Apart from this contribution to  the total potential of the
system one should also consider the energy density of the branes
that wrap a cycle over some of the compact dimensions. This
energy density is given by the volume spanned by the branes over
the compact directions times the tension of the branes. Then, the
energy density of our two brane system is  given by
\begin{equation}
E_0=E_1 + E_2 =T_p(A_1+A_2)=T_pA_T\,,
\label{etot}
\end{equation}
where $T_p$ denotes the tension of the D$p$-brane and is given by
$T_p=M_s^{p+1}/g_s(2\pi)^p$, being $g_s$ the string coupling and
$M_s$ the string mass. Also by $A_i$ we denote the ``volume''
spanned by the D$p_i$-brane along the compactified
directions\footnote{We will be considering for simplicity that in
the D$6$-brane case, $n^{(3)}_{(1)}=n^{(3)}_{(2)}=1$, and
$m^{(3)}_{(1)}=m^{(3)}_{(2)}=0$,  so that the branes are
separated in the $x_9$ direction.}
\begin{equation}\label{area}
A_i=(2\pi R)^{p-3}\sqrt{\left(\left(n^{(i)}_1\right)^2+
\left(m^{(i)}_1\right)^2
\right)\left(\left(n^{(i)}_2\right)^2+\left(m^{(i)}_2\right)^2\right)}.
\label{parallel}
\end{equation}

Note that if the configuration is supersymmetric the
energy~(\ref{etot}) does not vanish, as one would expect. This is
because we are not considering the total energy of the system. As
we explained in the previous subsection we need to introduce an
additional brane to cancel the R-R tadpoles. If we take into
account this additional brane the total energy will
cancel~(\ref{etot}) in the supersymmetric case.

At the end of the day, the total potential of the system is given by
\begin{equation}
V_T=E_0+V_{Dp}(\Delta\theta_I,Y)\,,
\label{total}
\end{equation}
where the exact form of $V_{Dp}(\Delta\theta_I,Y)$ depends on the
dimension of the branes that are present in the model, as it is
shown in~(\ref{pots}) and~(\ref{potf}). Recall that in~(\ref{total})
we consider that all the moduli are stabilized except for the
distance $Y$ between the branes, that will play the role of the
inflaton field.

\subsection{Brane recombination}
In this subsection we consider the evolution of the system
described before. Then we will be dealing with two D$5$ or
D$6$-branes with a four-dimensional intersection over the
non-compact dimensions and making two angles over the compact
dimensions (as shown in figure~\ref{fig:1}). The angles at which
the branes intersect each other, as can be seen in~(\ref{angle}),
are  given in terms of $(n_I^{(i)},m_I^{(i)})$, which denote the
number of times that the D$p_i$-brane wraps a cycle on the $I'$-th
torus. As we mentioned before this configuration is
supersymmetric when $\Delta\theta_1\pm \Delta\theta_2 =0$, and
preserves, in that case, $1/4$ of the supersymmetries.

Then, if $(a_i,b_i)$ denote the homology cycles of the torus,
the homology class of each D$p_i$-brane is given by the expression
\begin{equation}
[\Pi_{Dp_i}]=\left(n_1^{(i)}[a_1]+m_1^{(i)}[b_1]\right)\otimes
\left(n_2^{(i)}[a_2]+m_2^{(i)}[b_2]\right),
\end{equation}
being the total homology class of the configuration the sum of the homology
class of each~brane
\begin{equation}
[\Pi_{T}]=[\Pi_{Dp_1}]+[\Pi_{Dp_2}]\,.
\label{homc}
\end{equation}

When $\Delta\theta_1\pm \Delta\theta_2 \neq 0$ the configuration
is non supersymmetric and one of the complex scalars
in~(\ref{mdos}) becomes tachyonic triggering  the recombination
of the system. Since the branes  intersect at angles over the
compact space, the system will recombine into another  that wraps
a minimal volume supersymmetric cycle (that is, an special
lagrangian) over the compact dimensions. This cycle must have the
same total homology class than the initial
configuration,\footnote{ Remember that this has to be fulfilled in
order for the $RR$-tadpoles to be cancelled.} being this final
configuration a minimum of the tachyon potential.

As is shown in~(\ref{etot}) the energy of the initial
configuration is given in terms of the wrapping numbers of both
initial branes. Nevertheless, since this is  non supersymmetric
such configuration will not have a minimum energy or, in other
words, it will not saturate the BPS bound for the energy of a
system in the same homology class. The BPS bound for a
configuration of two D$p$-branes intersecting at two angles on a
$T^2\times T^2$ can be calculated in terms of the initial energy
of each brane~\cite{uranga}. The exact expression for this BPS
bound is given by
\begin{equation}
E_{BPS}=\sqrt{(E_1+E_2)^2-4E_1E_2\sin^2\frac{\Delta\theta}{2}}\,,
\label{en}
\end{equation}
where $\Delta\theta=\Delta\theta_1-\Delta\theta_2$, and $E_1$,
$E_2$, are just the energy of each brane in the initial
configuration. Note that when $\Delta\theta=0$ the configuration
is supersymmetric and so we recover the energy~(\ref{etot}) of the
initial configuration, as one would expect.

If we are consider configurations very close to the
supersymmetric case, then $\Delta\theta\ll 1$.
Expanding~(\ref{en}) around $\Delta\theta =0$ we get
\begin{equation}
E_f\simeq E_1+E_2-\frac{E_1E_2}{2(E_1+E_2)}(\Delta\theta)^2\,.
\end{equation}

The variation of the tachyonic potential is related to the
difference of the energy density between the initial and final
configurations~\cite{sen}, and in this case it is just given by
\begin{equation}
E_0-E_f\simeq \frac{E_1E_2}{2(E_1+E_2)}(\Delta\theta)^2\,.
\end{equation}

As we see, the difference  is proportional to the supersymmetry
breaking parameter $\Delta\theta$. This means that when
$\Delta\theta=0$  the difference of energy between both
configurations vanishes, as expected if one takes into account
that the supersymmetric case is stable and therefore there are no
tachyons.

The expression~(\ref{en}) gives us information about the energy of
the final configuration, but not about the cycle that it wraps.
In fact, in the general case, after the recombination of the
branes one will end up with just one brane with energy~(\ref{en})
that wraps a supersymmetric non-factorisable 2-cycle over the
compact space~\cite{uranga}. \pagebreak[3]

Nevertheless this is not the only possible final configuration.
As we will now show it is interesting to note that for a large
class of initial configurations there is also the possibility to
end, after the recombination of the system, with again branes
intersecting at two angles, but in a supersymmetric
configuration. This is an interesting possibility since
intersecting branes can have chiral matter at their intersections
and then are potentially useful to construct realistic models of
particle physics.

To analyze this possibility one has to take into account several
facts. The first thing is that for this recombination to occur,
one needs a final configuration that wraps also a factorisable
2-cycle with the same total homology class as the initial one,
which is given by~(\ref{homc}). That is, the final configuration
must satisfy
\begin{eqnarray}
[\Pi_{T}] &=& \left(n_{f1}^{(1)}[a_1]+m_{f1}^{(1)}[b_1]\right)\otimes
\left(n_{f2}^{(1)}[a_2]+m_{f2}^{(1)}[b_2]\right)+\nonumber\\[3pt]
&&+\left(n_{f1}^{(2)}[a_1]+
m_{f1}^{(2)}[b_1]\right)\otimes \left(n_{f2}^{(2)}[a_2]+
m_{f2}^{(2)}[b_2]\right),
\label{homcd}
\end{eqnarray}
where $(n_{fI}^{(i)},m_{fI}^{(i)})$ denotes the wrapping numbers
of the D$p_i$-brane over the homology cycles of the $I'$-th torus
in the final configuration. Written for each homology cycle the
eq.~(\ref{homcd}) tells us that the relation between the initial
and final wrapping numbers is the following
\begin{eqnarray}
n_1^{(1)}n_2^{(1)}+n_1^{(2)}n_2^{(2)} &=& n_{f1}^{(1)}n_{f2}^{(1)}+
n_{f1}^{(2)}n_{f2}^{(2)}\,,\nonumber\\[3pt]
n_1^{(1)}m_2^{(1)}+n_1^{(2)}m_2^{(2)} &=& n_{f1}^{(1)}m_{f2}^{(1)}+
n_{f1}^{(2)}m_{f2}^{(2)}\,,\nonumber\\[3pt]
m_1^{(1)}n_2^{(1)}+m_1^{(2)}n_2^{(2)} &=& m_{f1}^{(1)}n_{f2}^{(1)}+
m_{f1}^{(2)}n_{f2}^{(2)}\,,\nonumber\\[3pt]
m_1^{(1)}m_2^{(1)}+m_1^{(2)}m_2^{(2)} &=& m_{f1}^{(1)}m_{f2}^{(1)}+
m_{f1}^{(2)}m_{f2}^{(2)}\,.
\label{homcharge}
\end{eqnarray}

The second thing that we should take into account is that the
energy of the final configuration must saturate the BPS
bound~(\ref{en}). Nevertheless, if the final configuration has two
branes intersecting at two angles, then  its final energy must be
given by
\begin{eqnarray}
E_f &=& T_p(2\pi R)^{p-3}\Bigg[
\sqrt{\left(\left(n^{(1)}_{f1}\right)^2+\left(m^{(1)}_{f1}\right)^2
\right)\left(\left(n^{(1)}_{f2}\right)^2+
\left(m^{(1)}_{f2}\right)^2\right)}+\nonumber\\[3pt]
&&\hphantom{T_p(2\pi R)^{p-3}\Bigg[}
\! +\,\sqrt{\left(\left(n^{(2)}_{f1}\right)^2+\left(m^{(2)}_{f1}\right)^2
\right)\left(\left(n^{(2)}_{f2}\right)^2+
\left(m^{(2)}_{f2}\right)^2\right)}\Bigg]\,.
\label{enfin}
\end{eqnarray}

This means that in order for the final configuration to saturate
the BPS bound the expression for the energy~(\ref{enfin}) must be
equal to the expression~(\ref{en}). To see if that is possible it
is useful to rewrite~(\ref{en}) as
\begin{equation}
E_{BPS}=\sqrt{E_1^2+E_2^2+2E_1E_2\cos\Delta\theta}\,,
\label{endos}
\end{equation}
using that $\sin^2 {\Delta\theta}/{2}=\frac12(1-\cos\Delta\theta)$.
Now one can write $\cos\Delta\theta$ in terms of the wrapping numbers
of the initial configuration by means of~(\ref{angle}). With that and
also writing $E_1$ and $E_2$ as in~(\ref{parallel}) one finally
gets
\begin{eqnarray}
E_{BPS} &=& T_p(2\pi R)^{p-3}\Bigg[\left(n_1^{(1)}n_2^{(1)}+
n_1^{(2)}n_2^{(2)}\right)^2+
\left(m_1^{(1)}m_2^{(1)}+m_1^{(2)}m_2^{(2)}\right)^2 + \nonumber\\
&&\hphantom{T_p(2\pi R)^{p-3}\Biggl[}
\!+\, \left(n_1^{(1)}m_2^{(1)} +n_2^{(2)}m_1^{(2)}\right)^2
+\left(n_2^{(1)}m_1^{(1)}+n_1^{(2)}m_2^{(2)}\right)^2
\Bigg]^{1/2} \,.\qquad\qquad
\label{enhom}
\end{eqnarray}

Furthermore, as the final configuration must be supersymmetric
and we are not interested in final configurations with parallel
branes, the only non-trivial solution is given by configurations
that satisfy
\begin{eqnarray}
\left(n_{f1}^{(1)},m_{f1}^{(1)}\right)&=&
\left(n_{f2}^{(2)},m_{f2}^{(2)}\right)\,,\nonumber\\
\left(n_{f2}^{(1)},m_{f2}^{(1)}\right)&=&
\left(n_{f1}^{(2)},m_{f1}^{(2)}\right).
\label{con}
\end{eqnarray}
Replacing this into~(\ref{enfin}) the energy of the final
configuration will be given by
\begin{equation}
E_f=T_p(2\pi R)^{p-3}\sqrt{\left(2n^{(1)}_{f1}n^{(1)}_{f2}\right)^2
\!+\!\left(2n^{(1)}_{f1}m^{(1)}_{f2}\right)^2\!+\!
\left(2m^{(1)}_{f1}n^{(1)}_{f2}\right)^2
\!+\!\left(2m^{(1)}_{f1}m^{(1)}_{f2}\right)^2}\,.\quad
\label{efin}
\end{equation}

Now inserting the relations~(\ref{con}) for the final homological
charges in~(\ref{homcharge}) the conditions for the homological
charges become
\begin{eqnarray}
n_1^{(1)}n_2^{(1)}+n_1^{(2)}n_2^{(2)}&=&
2n_{f1}^{(1)}n_{f2}^{(1)}\,, \nonumber\\
n_1^{(1)}m_2^{(1)}+n_1^{(2)}m_2^{(2)}&=&
m_1^{(1)}n_2^{(1)}+m_1^{(2)}n_2^{(2)}
=n_{f1}^{(1)}m_{f2}^{(1)}+m_{f1}^{(1)}n_{f2}^{(1)}\,, \nonumber\\
m_1^{(1)}m_2^{(1)}+m_1^{(2)}m_2^{(2)}&=&
2m_{f1}^{(1)}m_{f2}^{(1)}\,.
\label{homchargedos}
\end{eqnarray}

Finally, just by using~(\ref{homchargedos}) and
comparing~(\ref{efin}) with~(\ref{enhom}) it is straight forward
to see that when
\begin{equation}
\mid n_{1}^{(2)}m_{2}^{(2)}-n_{2}^{(2)}m_{1}^{(2)}\mid
=\mid n_{f2}^{(1)}m_{f1}^{(1)}-n_{f1}^{(1)}m_{f2}^{(1)}\mid\,,
\label{cota}
\end{equation}
the energy of the final configuration saturates the BPS bound.
Note also that the condition~(\ref{cota}) is never zero since we
are considering that the initial configuration has always branes
intersecting at two angles. That means that we will only have
final configurations with branes at angles, not with parallel
ones.

Then we can conclude that if we  begin with an initial
configuration that satisfies
\begin{eqnarray}
\left(n_1^{(1)}n_2^{(1)}\!+\!n_1^{(2)}n_2^{(2)}\right)\!
\left(m_1^{(1)}m_2^{(1)}\!+\! m_1^{(2)}
m_2^{(2)}\right)\!&=&\!\left(n_1^{(1)}m_2^{(1)}
\!+\!n_2^{(2)}m_1^{(2)}\right)\!\left(m_1^{(1)}n_2^{(1)}\!+\!
n_1^{(2)}m_2^{(2)}\right),\nonumber\\
n_1^{(1)}m_2^{(1)}+n_1^{(2)}m_2^{(2)}&=&m_1^{(1)}n_2^{(1)}+m_1^{(2)}
n_2^{(2)}\,,
\end{eqnarray}
then the system will recombine into again two branes making two angles.

One can also  see that the final angle between the branes after
the recombination is just given by
\begin{equation}
|\tan\Delta\theta_f|=\frac{2|n_{1}^{(2)}m_{2}^{(2)}-
n_{2}^{(2)}m_{1}^{(2)}|}{n_1^{(1)}n_2^{(1)}+n_1^{(2)}n_2^{(2)}
+m_1^{(1)}m_2^{(1)}+m_1^{(2)}m_2^{(2)}}
\end{equation}
and also from here, since  $n_{1}^{(2)}m_{2}^{(2)}\neq
n_{2}^{(2)}m_{1}^{(2)}$, we see that in these cases we will not
have parallel branes after recombination, as we mentioned
previously.
\EPSFIGURE[r]{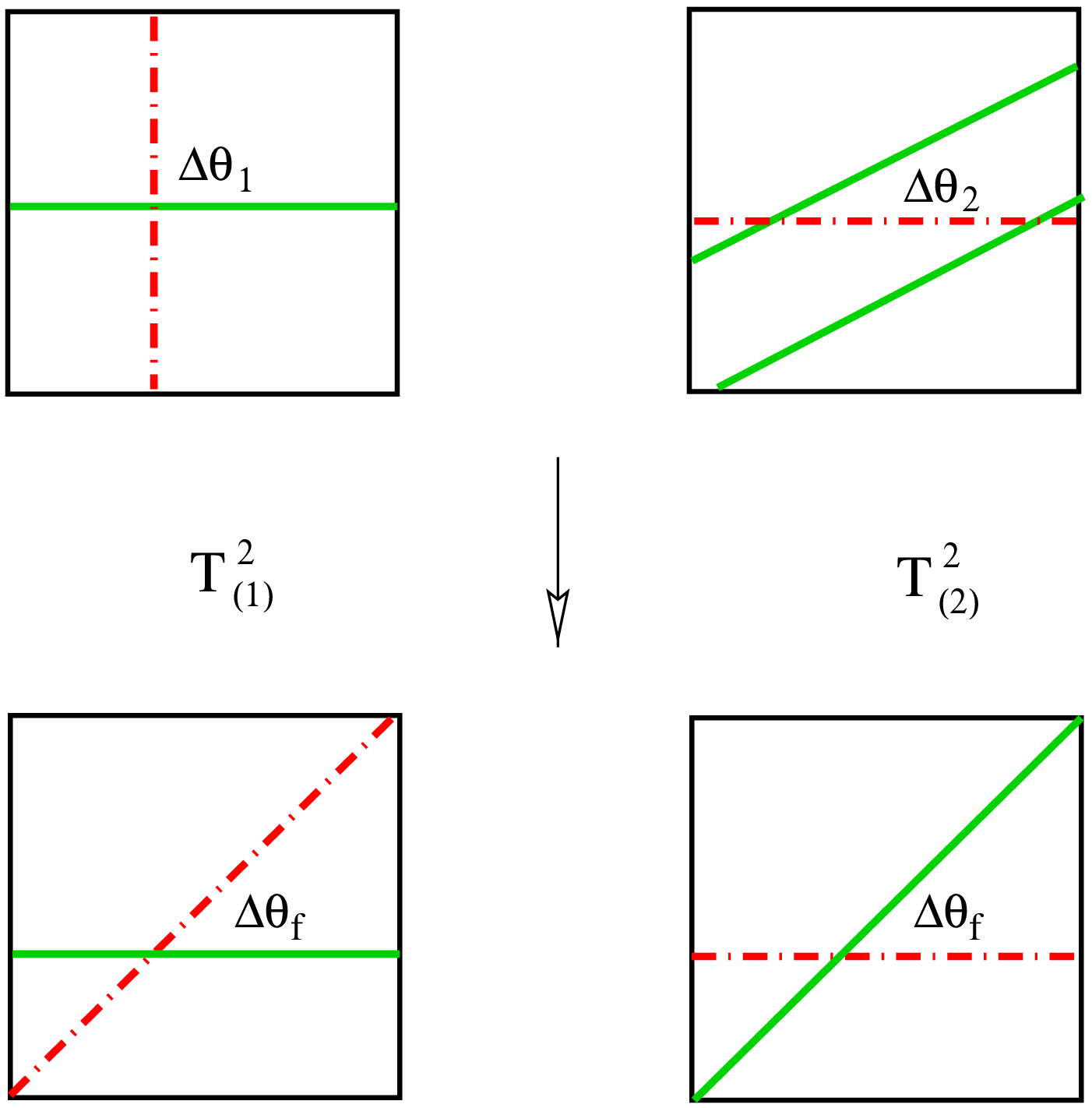, width=.361\textwidth}{Example of a concrete
configuration that recombines again into intersecting
branes.\label{fig:2}}

An example of such a recombination process is shown in
figure~\ref{fig:2}, in which we have taken two branes
intersecting in $T^2_{(1)}$ with wrapping numbers $(1,0)$ (green
solid line), $(0,1)$ (red dashed line), respectively, and in
$T^2_{(2)}$ with $(2,1)$, $(1,0)$. This brane configuration can
be recombined to give a configuration with again two branes at
angles but with wrapping  numbers $(1,0)$, $(1,1)$ on the first
torus and $(1,1)(1,0)$ on the second one, respectively. Note
that the angles that make the branes in both tori are the same so
the final configuration preserves $1/4$ of the supersymmetries.

Summarizing, we  have the following result: as we know, when the
tachyon appears, the branes will recombine into another brane
configuration that wraps a cycle that minimizes the energy. In
the general case the final brane configuration is only one brane
wrapping a non-factorisable 2-cycle on the compact space.
However, for some special  cases the final configuration  that
minimizes the energy contains  branes at angles but in a
supersymmetric configuration (that is, intersecting at the same
angle in both~tori).

We remark here again that  this fact  might have interesting
consequences from the point of view of building realistic models
from intersecting branes. Nevertheless we must point that in
order to obtain a chiral spectrum from these configurations one
should introduce extra objects in the model, such as for example
orientifolds in the spirit of the supersymmetric models described
in~\cite{uri}.

\section{Angled brane inflation}\label{sec:3}

In this section we discuss the inflationary parameters derived
from the potential between the branes at angles discussed in the
previous section. This potential will be valid as soon as the
distance between the branes is larger than the string scale,
$l_s$, and small compared with the radius of the torus, as the
potential was derived under these assumptions. When the distance
becomes too small, (see~(\ref{mdos})) the system will develop
tachyonic modes giving rise to a natural end of inflation via
extra fields (in this case tachyonic fields) as in the hybrid
inflationary models. This type of inflationary scenarios have
been discussed in string brane models for various
configurations~\cite{quei,queii,lusti,g-b}. This kind of mechanism
to end inflation (thorough the tachyonic field) seems to be
generic and it happens also in our present cases, as one might
have expected.

\subsection{Relevant inflationary parameters}
We consider now the relevant inflationary parameters that any
model of inflation should satisfy, and will study if the
constraints coming from those parameters are consistent with the
approximations we have done. That is, we will study if it is
possible to obtain a successful inflationary epoch within these
brane models intersecting at two angles.\pagebreak[3]

The four-dimensional effective action for the distance between the
branes, $Y$, can be written in the following form
\begin{equation}
S= \int{d^4x\sqrt{-g} \left[ \frac{1}{2}M_{P}^2\, \mathcal{R} -
\frac{1}{4}T_p \,A_T (\partial_\mu Y)^2 + V(Y)\right]},
\end{equation}
where $\mathcal{R}$ is the Ricci scalar in 4 dimensions,
$M_{P}^2= M_s^2 V_6\,g_s^{-2} = M_s^2(2\pi R
M_s)^6\,g_s^{-2}=2.4\times 10^{18}$\,GeV is the four-dimensional
Planck mass,  R is the radius of each square torus and $V_6$ is
the total volume of the compact dimensions. Then, the canonically
normalised field associated to the brane separation, $Y$, is
given by
\begin{equation}\label{Psi}
\Psi = Y \sqrt{\frac{T_p A_T}{2}} =
(Y M_{s}) M_s \sqrt{\sfrac{M_s^{p-3} A_T}{2\,g_s\,(2\pi)^p}}\,.
\end{equation}
Note that since $A_T\propto R^{p-3}$ the normalised field
$\Psi$ has dimensions of mass, as it should.

Let us now recall  the standard equations of motion of a
Friedmann-Robertson-Walker universe with a scalar field. These
are given by
\begin{eqnarray}
\ddot\Psi\,+\,3H\dot\Psi &=& -\frac{dV}{d\Psi}\,,\nonumber\\
H^2 &=& \frac{1}{3M_p^2}\left[V+\frac{\dot\Psi^2}{2}\right],
\end{eqnarray}
where $H$ is the Hubble parameter. The slow roll conditions
require that $|\ddot\Psi|\ll 3H|\dot\Psi|$ and $\dot\Psi^2\ll V$,
i.e.\ that  the friction and potential terms dominate. From these
conditions one can derive the two slow-roll parameters
\begin{equation}
\epsilon = \frac{M_P^2}{2}\left(\frac{V'}{ V}\right)^2\,,
\qquad \eta = M_P^2\frac{V''}{V}\,,
\end{equation}
which should be small in accordance with the field equations, so
that $|\epsilon| \ll 1$ and $|\eta| \ll 1$ for slow roll
inflation to occur. The primes above denote derivatives with
respect to the inflaton field $\Psi$.

The number of $e$-foldings occurring after the scales probed by
the COBE data leave the horizon can be computed as~\cite{liddle}

\begin{equation}\label{efolds}
N = \int{H dt} = \frac{1}{ M_{P}^2} \int_{\Psi_\mathrm{end}}^{\Psi_*}
\frac{V(\Psi)}{V'(\Psi)} d\Psi\,,
\end{equation}
where $\Psi_* \propto Y_*$ is the brane separation when the
primordial density perturbation exits the de Sitter horizon
during inflation. Since in this model the end of inflation is not
given by the cease of the slow-roll conditions in what follows,
we will assume that $Y_* \gg Y_\mathrm{end}$. This is a common
assumption in hybrid inflationary models and we will justify it
later.

The amplitude of the density perturbation when it re-enters the
horizon, as observed by Cosmic Microwave Background (CMB)
experiments is given by:
\begin{equation}\label{deltaH}
\delta_H = \frac{2}{5} \mathcal{P}_\mathcal{R}^{1/2} =
\frac{1}{5 \pi \sqrt{3}}\frac{V^{3/2}}{M_p^3\,V'}= 1.91\times 10^{-5}\,,
\end{equation}
where the value of $\delta_H$ is implied by the COBE results~\cite{COBE}.
\pagebreak[3]

The spectral index and its derivative\footnote{In order to study the
scale dependence of the spectrum, whatever its form is, one can
define an effective \emph{spectral index} $n(k)$ as $n(k)-1 \equiv
{d\,\ln{\mathcal{P}_\mathcal{R}}}/{d\,\ln{k}}$. This is
equivalent to the power-law behaviour that one assumes when defining
the spectral index as $\mathcal{P}_\mathcal{R}(k)\propto k^{n-1}$
over an interval of $k$ where $n(k)$ is constant. One can then work
$n(k)$ and its derivative by using the slow roll conditions defined
above.} can be expressed in terms of  the slow
roll parameters~\cite{liddle}, and they are given by
\begin{equation}\label{sindex}
n-1 = \frac{\partial\ln \mathcal{P}_\mathcal{R}}{\partial\ln k }
\simeq 2\eta - 6\epsilon\,,\qquad
\frac{dn}{d\ln k} \simeq 24 \epsilon^2 - 16\epsilon\eta +2\xi^2 \,.
\end{equation}
where $\xi^2\equiv M^2_P \frac{V'\,V'''}{V^2}$.

The gravitational wave spectrum can be calculated in a similar
form as for the scalar one, and the gravitational spectral index
$n_\mathrm{grav}$ is  given by

\begin{equation}\label{tensorpert}
n_\mathrm{grav}= \frac{d\ln\mathcal{P}_\mathrm{grav}(k)}{ d\ln k} =
-2 \epsilon \,.
\end{equation}
As we will see later, in  both the cases we will consider, the
variation of the spectral index and the gravitational
perturbations will be negligible, since the slow roll parameter,
$\epsilon$, turns out to be  too small, $\epsilon\leq 0.01$.

With these parameters we can now look separately at the two models of
branes intersecting  at two angles to determine if they can give us a
successful inflationary period.

\subsection{D$5$-brane model}
Consider now type-IIB string theory compactified on a squared,
factorisable, six-dimen\-sional torus $T^6$, as explained in
section~\ref{sec:2}. We now introduce two D$5$-branes which span
the four dimensional, $\mathbb R^{3,1}$,  non-compact dimensions
and intersect at two angles, $\Delta\theta_I$, between them in
two of the tori, as seen in figure~\ref{fig:1}. Intersecting
brane models of this type have been considered in~\cite{ibiii}.

The effective potential for the canonically normalised field
$\Psi$ (or $Y$) is obtained by combining the tension of the
branes and the interbrane potential energy (see
eq.~(\ref{total})). For the pair of D$5$-branes intersecting at
two angles in the compact space, and choosing an appropriate
cutoff in~(\ref{potf}), it is given by
\begin{equation}\label{D5pot}
V = T_5 A_T + \sfrac{M_s^4 F(\Delta\theta_I)}{(2\pi)^4}
\ln{\left[\frac{\Psi}{M_s} \right]},
\end{equation}
with $ T_5\,A_T = {(M_s^2\,A_T)M_s^4}/{g_s(2\pi)^5}$; the
index $I=1,2$ denotes the torus in which the branes make an angle
and from~(\ref{efe}) we have that,\footnote{Recall that we are
breaking slightly the supersymmetry so $\Delta\theta_1=\Delta
\theta_2 + \Delta\theta$ with $\Delta\theta \ll 1$.} $F\simeq
(\Delta\theta)^2$.

With this  potential~(\ref{D5pot}), the slow roll parameters become
\begin{equation}
\epsilon \simeq \frac{(2\pi)^2 g_s^2 F^2}{2 (M_s^2 A_T)^2}
\left(\frac{M_P}{\Psi}\right)^2, \qquad
\eta \simeq - \frac{2\pi g_s F}{(M_s^2 A_T)}
\left(\frac{M_P}{\Psi}\right)^2.
\end{equation}
Then it is clear that $|\epsilon|<|\eta|$, so that it would be
enough to consider  $\eta$ in the rest of the calculations.  So
in order for inflation to occur, we only need that the ratio of
$F$ to $M_s^2 A_T$ be small enough. That is, a successful
inflationary period will take place if
\begin{equation}\label{constraint}
\frac{F}{M_s^2 A_T} \ll 1\,.
\end{equation}
Notice that this requirement is very easy to achieve without
requiring a fine tuning of the parameters, since the volume $A_T$
can be done very large (see~(\ref{area})), and also $F$ can be made
very small. The number of e-folds can be calculated
from~(\ref{efolds}) to give
\begin{equation}\label{ene}
N \simeq \frac{(M_s^2 A_T)}{2g_s (2\pi)F}
\left(\frac{\Psi_*}{M_P}\right)^2,
\end{equation}
where $\Psi_*$ is proportional to the separation between the
branes when primordial density perturbation leaves the horizon
during inflation. Then, if~(\ref{constraint}) holds there would
be enough number of e-foldings for inflation to take place. In
terms of the number of e-folds, the slow roll parameters and the
spectral index can be written in the form
\begin{equation}
\epsilon \simeq \frac{1}{4\,N^2} \frac{\Psi^2_*}{M_P^2}\,,\qquad
\eta \simeq - \frac{1}{2\,N}\,,\qquad  n \simeq 1 - \frac{1}{N}\,;
\end{equation}
which is well within the present bounds from the CMB anisotropies for
$N \gtrsim 40$.

Now we have to impose the COBE normalisation, that is,
the amplitude of the scalar perturbations~(\ref{deltaH}) give us
\begin{equation}
\delta_H \simeq \frac{2^{1/2}\,(M_s^2 A_T)\,(M_s/M_P)^2\, N^{1/2}}
{5\pi\sqrt{3}\,(2\pi)^3\, g_s\, F^{1/2}}= 1.91\times 10^{-5}\,;
\end{equation}
and using the expression for $M_P$ we have from here that
\begin{equation}\label{efe5}
F^{1/2} = \frac{2^{1/2} N^{1/2} g_s (M_s^2A_T)}
{5\pi\sqrt{3}\,(2\pi)^3\,(1.91\times 10^{-5})(2\pi R M_s)^6}\,.
\end{equation}
The distance between the branes can be calculated
using~(\ref{Psi}),~(\ref{ene}) and~(\ref{efe5}). This gives us
\begin{equation}
M_s\,Y_* \simeq\frac{2\sqrt{2}\,g_s\,N}{5\pi\sqrt{3}
(1.91\times 10^{-5})\,V_6^{1/2}}\,.
\label{y}
\end{equation}
Note that it depends only on the overall compactification volume, $V_6$.

Now, if we take $N \sim 60$, with a weak  string coupling of $g_s
= 0.1$ and  a compactification radius of say, $2\pi R M_s = 25$,
we can compute the angle difference and the separation of the
branes at the beginning of inflation. From~(\ref{y}) one gets a
size for the separation between the branes of $M_s\,Y_* = 2.1 =
0.08 (2\pi R M_s)$, and then consistent with our approximations.

Also for the angular difference $\Delta\theta$ we get
\begin{equation}
\Delta\theta \sim 3.5\times 10^{-8}\, M_s^2 A_T\,,
\label{an}
\end{equation}
then the angle difference can be made larger using the volume
$A_T$, that is, making bigger the volume that the branes wrap
over the compactified dimensions. For example, for a volume of
order $M_s^2 A_T \sim 3\times 10^4$, the difference between the
angles will be $\Delta\theta\sim 10^{-3}$. The spectral index is
then $n\sim 0.98$ and its variation negligible as well as the
gravitational waves spectrum.  The string scale in this case is
$M_s ={M_P g_s}/{(2\pi R M_s)^3}\sim 1.54
\times 10^{13}$\,GeV, that is smaller than $M_P$ as one would expect since
we are considering weakly coupled string theory. These numbers
are quite common in models of inflation and we end up with a
successful inflationary scenario for an angular difference not so
fine tuned, and it can be enlarged  using $A_T$ (as it is shown
in~(\ref{an})). The end of inflation will be discussed in the next
section, however, as already mentioned, the appearance of the
tachyon will trigger the end of inflation as in any hybrid
inflationary model coming from non-stable brane configurations.

\subsection{D$6$-brane model}
Let us now consider type-IIA theory on $\mathbb R^{3,1}\times
T^6$ with the six torus factorisable, as discussed previously. We
now introduce two D$6$-branes which intersect at two angles in
the first two tori, but they are separated in the last torus by a
distance $Y$ (see figure~\ref{fig:1}). Intersecting models using
D$6$-branes have been studied widely in both
non supersymmetric~\cite{lustii,blu,blui,ibii} and
supersymmetric~\cite{uri} versions.

The effective potential for this model is given by~(\ref{total})
in terms of the distance between the branes $Y$.
Using~(\ref{Psi}) we can rewrite it in terms of the inflaton
field $\Psi$ and it becomes
\begin{equation}
V(Y,\Delta\theta_I) = T_6 A_T +
M_s^4\sfrac{V_6^{1/6}\,F(\Delta\theta_I)}{(2\pi)^3}\sqrt{\frac{2\pi
g_s}{M_s^3 A_T}}\,\left(\frac{\Psi}{M_s}\right).
\end{equation}
Here again $F \simeq (\Delta\theta)^2$ and $V_6^{1/6}= (2\pi R
M_s)$. The slow roll parameters become very simple in this case,
and in particular, notice that $\eta=0$. So we have,
\begin{equation}
\epsilon \simeq \frac{F^2\,(2\pi)^7\,g_s^3  V_6^{1/3}}
{2\,(M_s^3\,A_T)^3} \left(\frac{M_P}{M_s}\right)^2 .
\end{equation}
From here it is clear that in order to have an inflationary
period, the following condition has to be satisfied,
\begin{equation}\label{constraint6}
\frac{F^2}{(M_s^3\,A_T)^3} \ll 1\,,
\end{equation}
which again it is easy to get since the volume $A_T$ may be very
large and also $F$ very small. The number of e-folds can be
calculated using~(\ref{efolds}) to give
\begin{equation}
N \simeq \left( \frac{M_s}{M_P}\right)^2\frac{(M_s^3\,A_T)^{3/2}}
{(2\pi)^{7/2}g_s^{3/2}\,V_6^{1/6}\,F}\frac{\Psi_*}{M_s}\,.
\end{equation}
which will be big enough for inflation to occur as soon as the
constraint~(\ref{constraint6}) is satisfied. In terms of the
number of e-folds, the slow roll parameter and the spectral index
can be written in the form
\begin{equation}
\epsilon \simeq \frac{1}{2\,N^2}\frac{\Psi_*^2}{M_{P}^2}\,,\qquad
n \simeq 1 - \frac{3 \Psi_*^2}{N^2\,M_P^2} \sim 1\,.
\end{equation}
Since $\epsilon$ is very small, in this case the spectral index is
almost scale invariant and so its variation is going to be
negligible.

The amplitude of the scalar perturbations~(\ref{deltaH}) give us
\begin{equation}
\delta_H \simeq \frac{(M_s/M_P)^3 (M_s^3\,A_T)^2}
{5\pi\sqrt{3}(2\pi)^6\sqrt{2\pi} g_s^2  V_6^{1/6} F}\,.
\end{equation}
From here we can then compute the angle to be,
\begin{equation}
F = \frac{g_s\,(M_s^3\,A_T)^2}
{5\pi\sqrt{3}(2\pi)^6\sqrt{2\pi} (1.91 \times 10^{-5})
(2\pi R\,M_s)^{10}}\,.
\end{equation}
The distance between the branes is found to be
\begin{equation}
M_s\,Y_* \simeq \frac{\sqrt{2} N g_s}
{5\pi\sqrt{3} (1.91 \times 10^{-5}) V_6^{1/2}}\,.
\end{equation}
Note that this distance  depends only  on the square root of the
total compactified volume and not on the volume that the branes
wrap, as in the previous  D$5$-brane case (see eq.~(\ref{y})). In
fact it is easy to see that this is a general feature in models
with potentials of the form $V=T_p A_T+BY^{d}$, with
$d\in\mathbb{Z}$, or $V=T_p A_T + B\ln Y$.

In order to get some numbers,  consider again  $N\sim 60$  and  a
radius of  $2\pi R M_s \sim 25$, with $g_s$ as before, the
distance between the branes is $M_s\,Y_* = 2 = 0.08(2\pi R M_s)$
which is consistent with our approximations. Also for this values
one can easily compute the angular difference,
\begin{equation}
\Delta\theta \sim 1.3\times 10^{-9} \left(M_s^3A_T\right),
\end{equation}
then if we take a volume of order $M_s^3 A_T \sim 8\times 10^5$,
will give us $\Delta\theta \sim 10^{-3}$ and again it can be
increased using a larger volume. Note that for this values of the
parameters the string scale is of the same order as in the five
brane case. Also for this values $\epsilon\sim 10^{-12}$ and then
the gravitational perturbations will be again too small, so we do
not consider them here.

We have then seen that also in this simple case the potential
between the branes can give rise to an inflationary regime within
string motivated scenarios. The important issue which arises at
this point is the end of inflation, that will be triggered, as
was explained in previous sections, by a tachyonic field. We will
discuss this point briefly  in the next section.

\section{End of inflation}\label{sec:4}

The inflaton field $\Psi$ is proportional to the distance between
the branes $Y$, as can be seen in~(\ref{Psi}). Since the
potential of the system is attractive we will have that the
inflaton \mbox{field $\Psi$} rolls towards smaller values during
inflation. Nevertheless, inflation does not end when the slow roll
conditions ceased to be satisfied (i.e.\ $\epsilon\sim1$,
$\eta\sim 1$). The reason is that when the distance between the
branes becomes  small compared with the string scale, the
configuration develops tachyonic modes and many more degrees of
freedom apart from the interbrane distance become
relevant~\cite{seni}. This, as was first pointed out
in~\cite{quei}, provides a nice realization of a hybrid
inflationary scenario~\cite{linde}.

As explained in the first section this tachyonic fields will
trigger the recombination of the system. In general the final
configuration will be one brane wrapping a special lagrangian
cycle over the compact space. Nevertheless, as we mentioned
previously,\pagebreak[3] for \mbox{a large} class of initial
configurations the recombination of the system will be such that
it will allow for the final configuration to be also branes at
angles but in a supersymmetric configuration (that is,
intersecting at the same angle in both tori). This means that
also after inflation we will have a configuration with
intersecting branes and then suitable for building realistic
models of particle physics. Moreover as the volume that the
branes wrap in the initial configuration is not fixed by
inflationary constraints (that is, it is not fixed in order for
the relevant inflationary parameters to be satisfied), this means
that one can begin with a configuration that allows a suitable
recombination.

These systems develop a negative mode when the distance between
the branes in the compactified directions becomes smaller than a
critical value given in terms of the angles at which the branes
intersect. This critical value parametrises the end of inflation
and, as one can see from eqs.~(\ref{mdos}), it is given by
$M_sY_\mathrm{end}=\sqrt{(2\pi^2) \Delta\theta}$. Then when  $M_sY
\lesssim M_sY_\mathrm{end}$ tachyonic open strings connecting both
branes will appear. For our models,  $\Delta\theta\sim 10^{-3}$
and so, this value is given by $Y_\mathrm{end}\sim 0.14$.

It has been realised that the tachyon potential by itself does not
produce slow-roll inflation~\cite{quei}, but its importance arises
when inflation ends. In~\cite{rehe} it has been shown that the
roll along the tachyonic direction is very fast, and the initial
potential energy is quickly converted in gradient energy for the
tachyon field. The precise details of the reheating process are
out of the scope of the present letter,\footnote{For a more
detailed discussion on tachyonic reheating in this kind of
models, see for example~\cite{cline}.} however we can give a
crude approximation for the temperature of reheating. If we
assume that all of the initial vacuum energy of the branes is
converted into radiation, an estimate can be done by considering
$T_{RH}^4$ as the initial potential energy~\cite{quei}, giving
\begin{equation}
T_{RH} \sim  M_s \left[ \frac{\,M_s^{p-3}A_T}{
g_s (2\pi)^{p+1}}\right]^{1/4}.
\label{reh}
\end{equation}
If we introduce here the values for the volume spanned by the
branes used in the  cases considered, the D$5$ and D$6$-brane
models,  using $M_s\sim 10^{13}$\,GeV and $g_s\sim 0.1$ we get a
reheating temperature of order $T_{RH} \sim 10^{13}$\,GeV.

As we already mentioned, when the branes collide a tachyon
appears and the system recombine. Generically lower dimensional
D$(p-2k)$-branes can be formed during this process~\cite{seni,k}.
Since this  happens after inflation it is important to check
whether monopole-like objects or domain walls can be formed or
not, as the presence of such objects may destroy nucleosynthesis
or over close the universe. Nevertheless, as was pointed
in~\cite{st,tyei}, this does not seem to happen on these brane
configurations. The main reason is the fact that the size of the
compactified dimensions is much smaller that the Hubble radius
$H$. Then the only topological defects that can form
cosmologically by the Kibble mechanism whose production will not
be heavily suppressed are those of codimension two, where the
codimension should lie in the uncompactified space. That means
that only cosmic string-like defects are likely to be produced in
the four-dimensional space. The cosmic strings produced in this
way are in general D$(p-2)$-branes with $(p-3)$ dimensions over
the compact space, spanning the same compctified volume as the
former D$p$-brane, that we will denote by $A_p$ (as defined
in~(\ref{area})). Then, following the idea of~\cite{tyei} we can
calculate the tension of a cosmic string produced in this way to
be
\begin{equation}
\mu=T_{p-2}A_p=\frac{M_s^{p-1}}{(2\pi)^{p-2}g_s}A_p\,.
\label{mu}
\end{equation}
Gravitational interactions of the cosmic strings are given in
terms of the dimensionless parameter $G\mu$, where $G$ is the
Newton's constant. Using~(\ref{mu}) we can write this parameter~as
\begin{equation}
G\mu=\left(\frac{M_s}{M_P}\right)^2 \frac{M_s^{p-3}A_p}{(2\pi)^{p-2}g_s}=
\frac{g_sM_s^{p-3}A_p}{V_6(2\pi)^{p-2}}\,.
\end{equation}
This value must satisfy $G\mu\lesssim 10^{-6}$ in order not be in
conflict with present observations~\cite{COBE}. From here we can
estimate  an upper bound for the compactified ``volume'' of the
D$p$-branes $A_p$. This must satisfy
\begin{equation}
M_s^{p-3}A_p\lesssim 10^5 - 10^6\,.
\end{equation}
This value is within the range of volumes that we have used in
the previuos section. We should point out that this is only an
estimation and that a deeper analysis of the defect production
and reheating after inflation in the present scenario  would be
very interesting.

\section{Conclusions}\label{sec:5}

Along this paper we have considered interesting models of branes
intersecting at angles in a compact six-dimensional space. The
detailed configuration that we have studied is type-II string
theory on $\mathbb{R}^{3,1}\times T^6$ where the compact space is
a six-dimensional torus factorisable in three two-dimensional
tori. We introduced two D$p$-branes ($p=5,6$) in our
configuration in such a way that they intersected at angles on
two of the tori, but allowing them to be separated a distance $Y$
on the remaining one. These configurations are interesting since
they may be relevant in the construction of  realistic
--- intersecting at angles --- brane models (that is, models with a
spectrum close to the one of the standard model of particle
physics). Such models have been studied, for example,
in~\cite{lustii}--\cite{koko}. We have considered
two interesting cases, one in type IIA and the other one in
type-IIB string theory.

Like  in the case of branes intersecting at just one
angle~\cite{g-b,st}, we have found, in the two  cases considered,
that inflation is generic when the configuration is slightly
non supersymmetric. These models give a successful period of
inflation, consistent with observations. Moreover we have found a
remarkable feature when considering two angled configurations in
contrast with the one angled models. In fact, the configurations
we have considered, open up the possibility of interesting new
recombinations of the branes after inflation. In previous
models~\cite{g-b,st}, recombination at the end of inflation  gave
rise to the formation of only one brane wrapping some
supersymmetric cycle in the compact space. However, if one wants
to build realistic models of particle physics after inflation,
those models turned may not be very attractive,  since chiral
matter would not be present at the end of inflation.

We have shown how one can improve this situation by considering
branes intersecting at two angles. In this case, there is  the
possibility of ending up with only one brane wrapping a
supersymmetric non-factorisable 2-cycle in the compact space
after inflation. \pagebreak[3] Nevertheless there also exist the
possibility that the final configuration be that with two branes
intersecting at angles in the compact space, but in a
supersymmetric way, i.e.\ making the same angle in both tori.
This possibility is quite attractive since there are a number of
realistic models constructed from intersecting branes. There are
also some supersymmetric versions with intersecting branes
already in the literature~\cite{uri}. In any case, one should
note that independently of the type of recombination of the
branes after inflation we obtain, within these string D-brane
models, a successful inflationary epoch consistent with the
present observations.

Further investigation on the non-trivial recombinations of the
branes is worth to be done and also we must point out that this
recombinations may be useful in other aspects of string theory.
Another thing we want to remark is the fact that along this kind
of models one assumes that many moduli are fixed. It  would be
interesting to study how to stabilize these moduli in a
controlled way. A possibility may be the introduction of fluxes
in the configurations so one can use the techniques developed,
for example, in~\cite{flux,fluxi} to fix some of the moduli.
Finally, we mention again that a deeper investigation on the real
process of reheating and defect production after inflation is
definitely worth to be done.

\acknowledgments

We would like to thank J.~J.~Blanco--Pillado, F.~Marchesano,
R.~Rabad\'an, M.~M.~Sheikh-Jabbari, N.~V. Suryanarayana,
M.~Trapletti and A.~Uranga  for many useful discussions, and
M.~E.~Angulo and L.E.~Velasco-Sevilla for collaboration at early
stages of this work. We would also like to thank specially
F.~Quevedo for many discussions and for the critical reading of
the manuscript. The work of I.Z.C. is supported by CONACyT,
Mexico, Trinity College (Cambridge) and COT (Cambridge Overseas
Trust). The work of M.~G.-R. is supported by Fundaci\'on Ram\'on
Areces.

\end{document}